\documentstyle[12pt]{article}
\begin{document}

\title{ The decay $H \to \gamma\gamma$ in Multi-Higgs doublet models}
\author{   R. Martinez, J.-Alexis Rodriguez and M. Vargas\\
 Dpto. de F\'{\i}sica, Universidad Nacional de Colombia}
\date{}
\maketitle

\begin{abstract}

We study the dominant decays of the lightest Higgs boson 
in models with 2 and 3 Higgs doublets, for the case when
its couplings to fermions are absent at tree-level. 
It is found that the branching ratio 
for the decay $H\to \gamma\gamma$ is above the 
one into fermion pairs, which is evaluated also at the 1-loop level.
\end{abstract}

\pagebreak

The search for the Higgs boson of the standard model (SM) \cite{sma},
is the most important test of the symmetry breaking and mass
generation of the theory. 
Current limits on the SM Higgs mass coming  from direct searches for the Higgs boson, specifically from the study of the reaction $e^+ e^- \to Z \to (Z^* \to ff) h$  at LEPI. The combined limit of the four experiments on the Higgs mass is $m_h >65.4$ GeV \cite{blondel}. At LEPII with the total energy $\sqrt{s}=130-200$ GeV, the dominant Higgs production process is $e^+e^- \to hZ$, where the final state particles in the analysed Higgs boson channels are $e^+ e^- \to (Z \to qq , bb, \nu \nu, \tau \tau, e^+e^- ,\mu \mu) (h \to bb, \tau \tau)$. Combined limit  of the four experiments with $\sqrt{s}=195.6$ GeV gives $m_h \geq 102.6$ GeV at $95 \%$ C.L.. LEPII running with the total energy $200$ GeV will be able to discover standard Higgs boson with a mass up to $107$ GeV \cite{lepa}. 

On the other hand, indirect bound on the Higgs mass can be obtained from
precision electroweak measurements. Although the sensitivity to the Higgs boson mass through radiative corrections is only logarithmic, the increasing precision in the measurement of electroweak observables allows to derive constraints on $m_h$, around of $m_h=71^{+75}_{-42} \pm 5$ GeV \cite{hollik}. Other constraints coming from tree level unitarity in $W_L-W_L$ scattering, $m_h \leq 1$ TeV \cite{quigg}, 
validity of perturbation theory, $m_h \leq 930$ GeV \cite{hollik}, and the analisys done of vacuum stability, $m_h \geq 120$ GeV \cite{24}.  On the other hand, the search techniques
for intermediate and heavy Higgs boson are known, and their implementation 
requires the next generation of hadron colliders LHC \cite{all}. One of the most important reactions for the search for the Higgs boson at LHC is $pp \to (h \to \gamma \gamma)$ which is the most promising one for the search in the region $100 \leq m_h \leq 140$ GeV \cite{lepa}. 

In this paper we shall study the Higgs sector for
two particular models, which contain 2 and 3 Higgs doublets respectively,
for the case when some Higgs boson couples only to gauge bosons but not
to fermions, at tree-level. In this case the
decay into $\gamma\gamma$ is expected to dominate for the
lower part of the intermediate mass region, 
however the decays into fermion pairs can be generated also at the
1-loop level, and one needs to include their contribution into the total
width, in order to know the precise values of branching ratios,
which to our knowledge has not been done in the literature \cite{hgams}.

 We shall discuss first the Lagrangian for the 
model with two doublets,then
the dominant branching ratios for the intermediate mass
range are evaluated, and we also discuss
the limits on the Higgs mass that can be obtained from the LEP data.
 Later on, it will be explained how to
obtain the corresponding results for the 3-Higgs doublet model.

The extension of the SM with two Higgs doublets has been studied in
great detail before \cite{twohdm}. 
In terms of components the two doublets are written as:
$\Phi_1=(\phi^+_1,\phi^0_1)^T$, $\Phi_2=(\phi^+_2, \phi^0_2)^T$.
It happens that the Yukawa coupling of
the Higgses with the fermions can be choosen in two ways,
 usually denoted as models I and II. In model I, one doublet
is used to generate masses for both the U- and D-type quarks, 
 whereas in model II one doublet generates the 
masses of the U-type quarks and the second one generates the masses of
the D-type quarks. We shall consider here only model I.

After diagonalizing the general Higgs potential, 
one gets the scalar mass
eigenstates, which include one charged pair ($H^\pm$), two CP-even scalars 
($h^0$, $H^0$, with $m_H > m_h$), and one CP-odd scalar ($A^0$).
Thus, the free parameters are the scalar masses, the mixing angle $\alpha$
and the ratio of vev's $\tan\beta=v_2/v_1$. 
The mass eigenstates can be written in terms of the components of the
Higgs doublets; for the CP-even scalars for example, one has:
\begin{equation}
 H^0  = 2^{1/2}[ (Re \phi^0_1-v_1)\cos\alpha+ 
(Re \phi^0_2-v_2)\sin\alpha ] 
\end{equation}
\begin{equation}
 h^0  = 2^{1/2}[- (Re \phi^0_1-v_1)\sin\alpha+ 
(Re \phi^0_2-v_2)\cos\alpha] 
\end{equation}

The interaction of fermions with Higgs bosons in model I are given by
the Yukawa Lagrangian,
\begin{eqnarray}
L = &-& {g \over 2m_W\sin\beta}
 [{\bar D}M_D D+{\bar U}M_U U] (H^0\sin\alpha+h^0\cos\alpha) \nonumber\\
  & + &{ig\cot\beta \over 2m_W}
 [-{\bar D}M_D\gamma_5 D+{\bar U}M_U\gamma_5 U] A^0 \nonumber\\
  &  + & {g\cot\beta \over 2^{3/2}m_W} 
  ( H^+ {\bar U} [ M_U K P_L - K M_D P_R] D + h.c.)  
\end{eqnarray}
where $P_{R,L}=(1\pm \gamma_5)/2$, $M_{U,D}$ are the diagonal mass
matrices of the U- and D-type quarks, $K$ is the Kobayashi-Maskawa
mixing matrix. It happens for this type of models that the important
coupling $h^0VV$ ($V=Z, W$) is proportional to
$\sin(\beta-\alpha)$.
Moreover, if the mixing angle takes the value $\alpha=\pi/2$, then
there is no mixing among $H^0$ and $h^0$, and one has that 
$h^0$ interact only with the gauge bosons, with the coupling
proportional to $\sin(\beta-\pi/2)=-\cos\beta$.

The dominant decays of $h^0$ in the mass range $m_h > 2m_Z$ 
are into $WW$, $ZZ$; whereas for the intermediate mass range
(80 GeV $< m_h<2m_Z$) the allowed decays are into $\gamma\gamma$, $Z\gamma$ ,
$WW^*$, $ZZ^*$ , which will compete also with the decays into
fermion pairs, generated at the 1-loop level.

The decay width into photon pairs can be written as:
\begin{equation}
 \Gamma(h^0 \to \gamma\gamma)= \cos^2{\beta}
\Gamma^W_{sm}(h^0 \to \gamma\gamma), 
\end{equation}
where $\Gamma^W_{sm}$ denotes the W-loop contribution to the decay width
of the SM Higgs boson; the decay width for $h^0\to Z+\gamma$
has also the same form, and it will be bellow the one for 
$h^0\to \gamma\gamma$, as in the SM case, thus we shall not discuss it
further here.
Similarly, we find that the decays into $WW^*$ and $ZZ^*$ can be written
in the same form, namely \cite{all}
\begin{equation}
\Gamma (h^0\to VV^*)= \cos^2\beta\ \Gamma(\phi^0_{sm} \to VV^*)
\end{equation}
Finally, the expression for the decay width into fermion pairs
($h^0\to f \bar f$), that results after one evaluates 
the 1-loop amplitude is written as follows:
\begin{equation}
 \Gamma(h^0\to f \bar f)={ G_F \alpha^2 \pi \over 2\sqrt{2}\sin^4\theta_W}
 m_h m^2_f \cos^2\beta F(m_h,m_i,m_W) 
\end{equation}
where $m_i$ is the mass of the fermion that enters in the loop,
and $F(m_h,m_i,m_W)$ is a function that arises from the
loop integration, which is written as follows,
\begin{equation} F =[|F_1|^2+|F_2|^2] |K_if|^2, \end{equation}
where,
\begin{eqnarray}
F_1 &= &4m^2_W C_{12}+m^2_h(C_{0}-C_{12}+C_{23}-C_{11})+m^2_i(C_{12}-C_0), \nonumber  \\
 F_2&=& 4m^2_W(C_0-C_{11}+2 C_{12})-m^2_h(-C_0+C_{11}+C_{12}+C_{23}) \nonumber \\
&& + m^2_i(2C_{11}-C_0), 
\end{eqnarray}
where $C_{ij}=C_{ij}(m_h,m_i,m_W)$ can be written 
in terms of the
scalar integral $C_0$, as discussed in \cite{scalin}. 
From this expression one notices that the width is
again proportional to $m_f$, which will suppress the width, thus only the
heaviest fermions will contribute significantly. This result can 
be understood easily if one follows the chirality in the graphs,
which need a mass insertion to be different from zero. In the following
analysis we include only the contribution from the top quark to
the loop, which is the dominant one.
The resulting branching ratios are presented in fig. 1, where
one can see that the decay into a photon pair dominates for
mass values of the Higgs up to about $m_Z$.

On the other hand, 
if one considers a model which allows Higgs-fermion couplings,
and one assumes that there are no 
Flavour Changing Neutral Currents (FCNC) 
mediated by the neutral Higgs bosons, 
then the couplings $h^0 ff$ are proportional to $m_f$, and then the
production rates will be highly suppressed; whereas
if one allows for the presence of FCNC, then the 
Higgs-fermions couplings are not neccessarly proportional to the
fermion masses \cite{mymod}, and the cross-section can be large. 

One can use the experimental results to
constrain the mass of the Higgs boson for this kind of models.
This can be done using the LEP bound 
$BR(Z\to \nu\nu \gamma\gamma)< 10^{-6}$ \cite{lephg}, which
can be written for this model as,
\begin{equation}
BR(Z\to \nu\nu \gamma\gamma)=BR(Z\to \nu\nu + h^0)
               BR(h^0\to  \gamma\gamma),   \end {equation}
which depends only on $m_h$ and $\tan \beta$.
 
Fig. 2 shows the excluded region in the plane $\tan \beta-m_h$, obtained from
the previous equation. It is important to point out that this is the first
bound that is obtained for a model of this type, which eventhough 
is valid only for an specific value of $\alpha$, 
it does not depends on the remmaining parameters 
of the general Higgs potential. For low values of $\tan \beta$ the
limit on the Higgs mass is $m_h>91 $ GeV, which is similar to
the one obtained for the SM Higgs.

Finally, we have also evaluated the branching ratios of Higgs bosons
within the context of
a  model with 3-Higgs doublets, where 2 of them behave like the doublets
of model II in the 2-doublet case. The third doublet couples
only to vector bosons, and we find again that the dominant decay 
in the intermediate mass range, is into photon pairs.

The scalar potential for this model, which allows for the existence of
one CP-even Higgs boson that does not couples at fermions, is the
following,
\begin{equation} V= V(\Phi_1,\Phi_2)+V(\Phi_3) \end{equation}
where $V(\Phi_1,\Phi_2)$ is the two-Higgs doublet model
\cite{twohdm}, whereas $\Phi_3$ contains the Higgs
scalars that do not mix with the other Higgs bosons;
each of the doublets can be written as 
$\Phi_i= [ \phi^+_i, h_i+v_i+ i \eta_i ]$.
Then, $h_3$ can be choosen as the light Higgs that only couples to
gauge bosons, whose coupling is $g_{h_3VV}=\sin\gamma g_{\phi^0_{sm}VV}$
(where $\tan\gamma=v_3/(v^2_1+v^2_2+v^2_3)^{1/2}$).

 We have derived all the relevant Feynman rules of the model,
needed to evaluate the decay widths at tree-level and 1-loop,
and the final results is that
all the branching ratios of $h_3$ can be obtained from
the ones obtained previously for the two Higgs
doublets model I, just by replacing $-\cos\beta \to \sin\gamma$. Thus, the
previous limits on the Higgs mass apply also for this scalar.
Models with N-doublets have been analyzed in the literature too \cite{nhix},
and it is possible to translate our limits for such models by taking the
appropriate limit.

In summary, we find that the scalar sectors studied 
in this paper have an interesting phenomenology in their own. And,
 it is possible to use the LEP results to put limits on the Higgs
mass, which are comparable to the ones obtained for the SM Higgs
for low values of $\tan\beta$, namely $m_h> 91$ GeV.

 At $e^+e^-$ machines with TeV CM-energies, it will be 
possible to produce the Higgs boson of these models 
by WW fusion, and also in association with Z.
 The production of a Higgs by photon-photon fusion, 
could be also important too, unfortunately 
we found that at LEP energies the event rate is far bellow detectability.
 On the other hand, the phenomenological consequences of these models at 
hadron colliders are also interesting. Because of 
the absence of Higgs-fermions couplings, it will not be possible to
produce the Higgs in association with top pairs, and neither by
gluon fusion (which will occur only at the 2-loop level), which
have proved to be usefull for the SM case.
Thus, the main production mechanism will be in association with W/Z,
however since the decays into gauge bosons will dominate, its detection
could be feasible.
Clearly, a detailed study is needed in order to
determine the detection feasibilities of these modes, 
which is beyond the scope of this work.

We acknowledge discussions with L. Diaz-Cruz. This work has been finantialy supported by COLCIENCIAS(Colombia) and CONACyT (Mexico).


\vspace{2cm}

\begin{center}
{\bf Figure Caption}
\end{center}

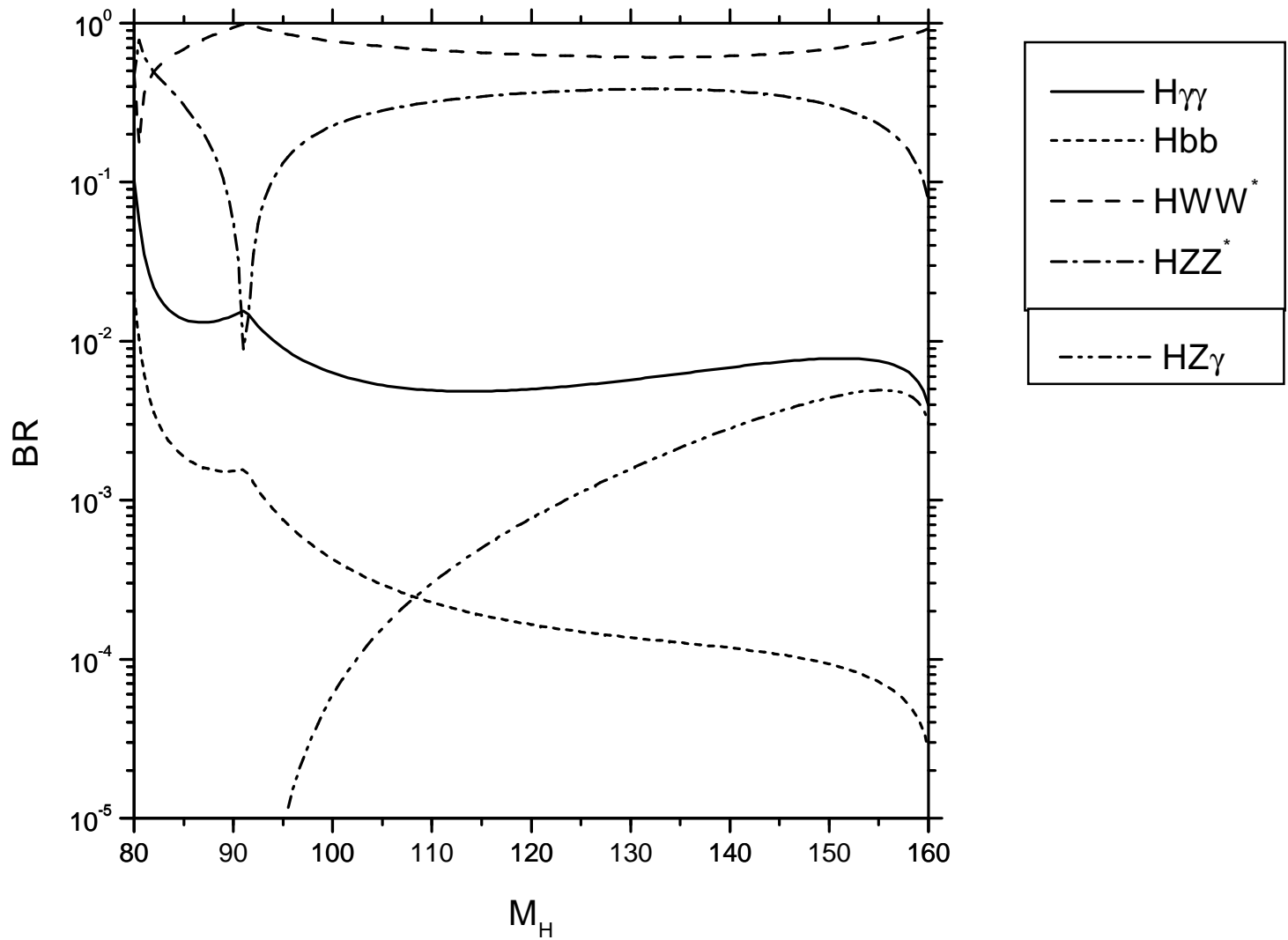
\noindent Figure 1: Branching ratios for the decay of the Higgs boson; $h\to \gamma\gamma$: solid , $h\to b\bar b$: dot-dash, $h\to WW^*$: dashes, $h\to ZZ^*$: dots.

\vspace{3cm}

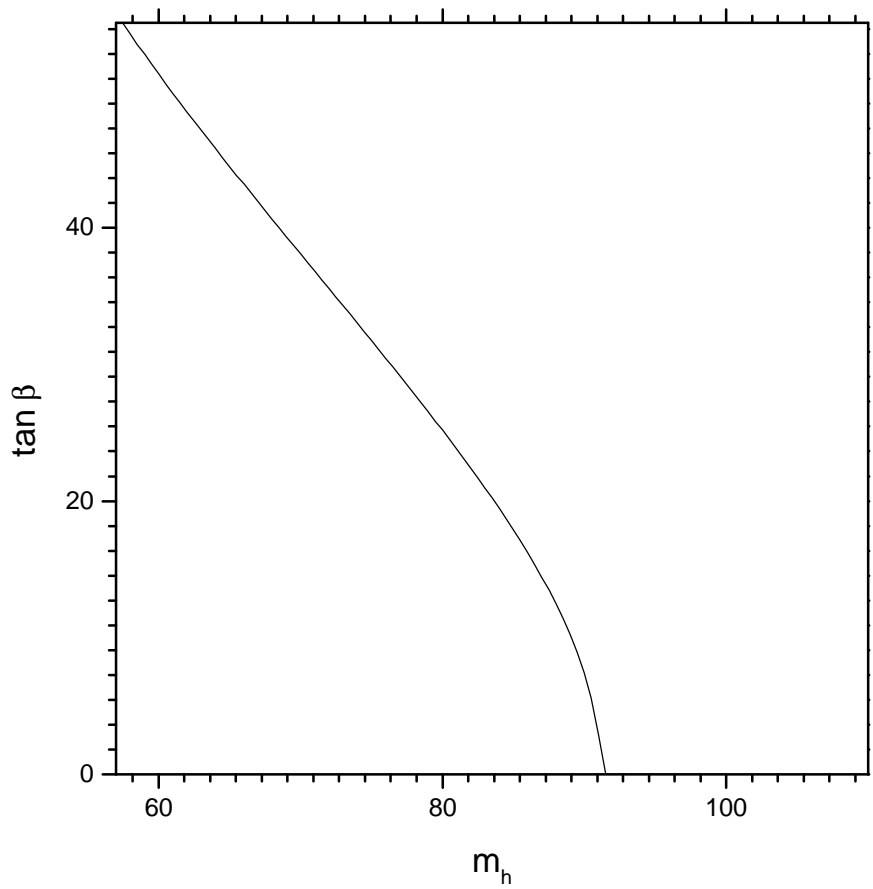
\noindent Figure 2: Regions in the plane $\tan\beta-m_h$ excluded by the
LEP results (shaded), for the models discussed in the text.

\end{document}